\documentclass[twocolumn,aps,showpacs]{revtex4}
\usepackage{graphicx,amssymb,amsmath}

\begin{document}

\title{Thermoelectric Power in the Double Exchange Model}

%\author{Beom Hyun Kim,
%   Unjong Yu\footnote{Current address: 
%   Theoretical Physics III, Center for Electronic 
%   Correlations and Magnetism, Institute for Physics, University of 
%   Augsburg, D-86135 Augsburg, Germany }
%   Kyoo Kim, and
%  B. I. Min\footnote{Corresponding author: {\sf bimin@physics.postech.ac.kr}}}
%\affiliation{PCTP, Department of Physics, Pohang University of Science and
%        Technology, Pohang 790-784, Korea}

\author{Beom Hyun Kim$^1$, Unjon Yu$^2$, Kyoo Kim$^1$, and 
 B. I. Min$^1$\footnote{Corresponding author: {\sf bimin@physics.postech.ac.kr}}}
\affiliation{$^1$PCTP, Department of Physics, Pohang University of Science and
	Technology, Pohang 790-784, Korea \\
	 $^2$Theoretical Physics III, Center for Electronic
       Correlations and Magnetism, Institute for Physics, University of 
       Augsburg, D-86135 Augsburg, Germany}
\date{\today}

\begin{abstract}
Employing the Monte-Carlo method and the exact diagonalization, we have
investigated the temperature dependence of the thermoelectric power (TEP)
for the double exchange model in the dilute carrier concentration limit.
We have found that the TEP follows the Heikes formula
in the high temperature regime, whereas,
in the intermediate temperature regime, 
the TEP is suppressed by the exchange coupling between itinerant
electrons and local spins. 
In the low temperature regime, 
the TEP exhibits an anomalous peak and dip feature
near the magnetic transition temperature $T_C$ 
which can be understood based on the magnetic polaron state. 
We have also found that the TEP, in the presence of the magnetic field,
shows the positive magnetothermoelectric power near $T_C$.
\end{abstract} 

%\keywords{Thermoelectric Power;Double Exchange;Magnetothermoelectric Effect;
%Magnetic Polaron}

\maketitle
%==============================================================================

%\section{Introduction}

The strong coupling between the electrical transport and the magnetism 
appears in the systems which show the transition from the paramagnetic (PM) 
insulator to the ferromagnetic (FM) metal. Colossal magnetoresistance (CMR)
doped manganese oxides \cite{Nagaev}, dilute magnetic semiconductors
\cite{Ohno,Dietl}, and EuB$_6$ \cite{Guy,Pasch} are typical examples. 
Simultaneous FM and metallic transitions
have been well described by the double exchange model
in which
the itinerant electrons are coupled to the local spins 
through the Hund-type exchange interaction.\cite{Zener,Arovas,Letf,Dagotto}

In these systems, the thermoelectric power (TEP) as well
as the electrical conductivity shows an anomalous behavior due to
a strong correlation between conduction electrons and local spins. 
For example, 
depending on the doping ratio, applied magnetic field, or temperature,
doped perovskite manganese oxides exhibit intriguing TEP behaviors such as
the sign change, the appearance of a large peak in the FM phase, 
and the magnetothermoelectric effect.\cite{Asam,Hundl,Mandal,Beb}
The TEP would be affected by various interactions such as electron-electron,
electron-phonon, Jahn-Teller interaction, etc. 
The essential feature of the TEP in doped perovskite manganese oxides,
however, can be the explained by the double exchange mechanism.

It has been well known that the double exchange model in the 
dilute concentration limit produces the magnetic polaron.
The magnetic polaron is a composite quasi-particle of a charge carrier and 
the magnetic polarization field of local magnetic moments induced 
by the carrier.\cite{deGennes,Mott}
The magnetic polaron effect enhances the effective
mass of the carrier and can localize carriers.
Thus the magnetic polaron state is featured by FM clusters embedded in 
the paramagnetic background and by the insulating behavior 
in the resistivity.\cite{yu05}

In this study, we have examined the TEP behavior for
the double exchange model in the dilute concentration limit.
By using the Kubo's formalism,
we have obtained the temperature dependent TEP and investigated its behavior
with respect to 
the exchange coupling strength. Further, we have demonstrated
the field dependence of the TEP near the magnetic transition temperature $T_C$.

%\section{Theoretical formalism}

The double exchange model 
which includes the strong exchange interaction
between itinerant electrons and local spins has the following 
Hamiltonian form:
\begin{eqnarray}
\label{Hamiltonian}
H &=& - t \sum_{\langle i,j \rangle \sigma} 
            \left( c^{\dagger}_{i \sigma}
            c^{}_{j \sigma} + \text{h.c.} \right) 
      - J_{H} \sum_{i} \vec{\sigma}_i \cdot \vec{S}_i,
\end{eqnarray}
where $\vec{S}_i$ and $\vec{\sigma}_i$ denote 
the spin operators of the local and the itinerant electron 
at the $i$-th site, respectively. 
$c^{}_{i \sigma}$ ($c^{\dagger}_{i \sigma}$)
is the annihilation (creation) operator of the electron 
with the $\sigma$ spin state at the $i$-th site, $t$ is the hopping parameter,
and $J_{H}$ represents the exchange coupling between local spins and
electrons.
For simplicity, we have considered, in this study, 
a two-dimensional ($2D$) square lattice system with Ising-type local spins 
($S_z=\pm \frac{1}{2}$).
Previous studies show that the simulation
results of the double exchange model do not depend much on
the dimensionality and lattice type.\cite{Yunoki,Dagotto2}
We have neglected the electron-electron interaction.
The electron-electron interaction can give rise to the 
important effect on the electrical and thermal transport properties. 
To avoid its effect, we have considered a system with 
very low carrier concentration.

In order to calculate the TEP for the double exchange model, 
we have employed the Kubo's formalism of the transport coefficients 
for a system in the presence of both electric field and temperature
gradient.\cite{Lutt,Zeml,Shastry}
The thermoelectric power is given by
\begin{equation}
\label{Thermo}
 S = \frac{1}{T} \frac{M_{12}}{M_{11}} + \frac{\mu}{|e| T},
\end{equation}
where $e$ is the electron charge and $\mu$ is the chemical potential.
The transport coefficients $M_{ij}$ are defined by
\begin{multline}
 M_{ij} = \lim_{\omega \rightarrow 0} \Bigl{\{} D_{ij} \delta(\omega) \\
          + \frac{1}{N \hbar} \frac{1-e^{-\beta \hbar \omega}}{2\omega}
    \int_{-\infty}^{\infty} dt e^{i \omega t} 
    \langle \hat{J}_{i}(t) \hat{J}_{j}(0) \rangle  \Bigr{\}},
\end{multline}
where $N$ is the size of lattice, $D_{11}$ and $D_{12}$ are the general
stiffnesses of the charge and energy, and 
$\hat{J}_{1}$ and $\hat{J}_{2}$ represent their current operators, respectively.

Using the Monte-Carlo (MC) method combined with the exact diagonalization (ED),
we have calculated the thermal average of the TEP.
For the description of thermal contribution of local spins, 
we have used the MC method based 
on the standard Metropolis algorithm \cite{Metro} with
the periodic boundary condition.
The electronic energy and states are obtained through the ED method.\cite{ed}

%\section{Results and discussions}

\begin{figure}[t]
\includegraphics[width=7.8cm]{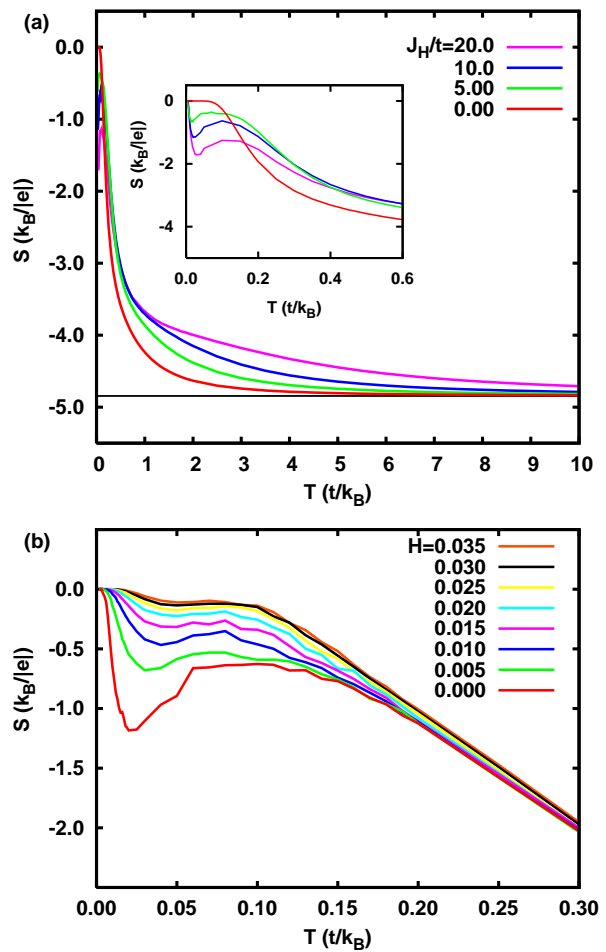}
\caption{(color online). (a) The temperature dependence of the thermoelectric
power $S(T)$ for different $J_H$ parameters in the double exchange model.
The thin horizontal solid line which crosses $S=-4.8442~k_B/|e|$ 
represents the value of $S$ calculated by the Heikes formula.
The inset in (a) presents the low temperature behaviors of $S(T)$.
(b) The magnetic field dependence of $S(T)$ for $J_H/t=10.0$.
For all cases, $n$ (electron density) is fixed by $n=0.0156$
(one electron in an $8 \times 8$ lattice).  }
\label{fig1}
\end{figure}

In the high temperature limit, 
the TEP for the double exchange model becomes nearly constant 
and simply proportional to the entropy per carrier.
It is because the first term in Eq.~\ref{Thermo} becomes zero at high
temperature and only the second term contributes to the TEP.
Then the TEP is expressed by the following Heikes formula:\cite{Chai}
\begin{equation}
\label{Heikes}
 S(T \rightarrow \infty) \rightarrow  \frac{\mu} {|e| T} 
 = \frac {s} {|e|}
 = -\frac {k_B} {|e|} \ln \frac {2-n} {n},
\end{equation}
where $s$ is the entropy per electron, and 
$n$ is the electron density.
This feature is well described in Fig.~\ref{fig1}(a). 
Regardless of the magnitude of $J_H$, 
all the curves of the TEP converge to the specific horizontal line.
This asymptotic line in Fig.~\ref{fig1}(a) represents the value of the TEP
($S=-4.8442~k_B/|e|$) for the case of dilute concentration $n=0.0156$.

%==========================================================
%==========================================================

In the intermediate temperature regime ($0.5<T<10.0$), 
the TEP exhibits different behaviors depending on the strength of $J_H$. 
As shown in Fig.~\ref{fig1}(a), 
the value of TEP is the largest for $J_H=0.0$,
and  becomes reduced with increasing $J_H$.
It is because the strong exchange coupling between an itinerant electron and 
local spin brings down the spin entropy of conduction electrons and 
inhibits the heat current.

In the low temperature limit ($T<0.5$), 
the TEP for the double exchange model shows interesting behavior 
for finite $J_H$. 
The inset in Fig.~\ref{fig1}(a) reveals that
a peak and dip feature occurs in the TEP curve for $J_H \ne 0$. 
Note that the TEP for $J_H=0.0$ converges monotonically to zero.\cite{FS}
Transport properties of a metallic system are well described 
by the semi-classical Boltzmann transport theory \cite{Boltz},
which yields the $T$-linear TEP  behavior at low temperature.
On the other hand, the TEP for an insulating system 
is known \cite{Girvin} to be proportional to $1/T$. 
Thus it is inferred from the inset in Fig.~\ref{fig1}(a)
that an insulating phase appears above the dip position temperature
$T_D$, but disappears with decreasing temperature below $T_D$. 

As mentioned above, the double exchange model 
in the dilute carrier concentration limit
is illustrated by the concept of the magnetic polaron.\cite{yu05}
Below the peak position temperature $T_P$, 
the strong exchange coupling produces the composite quasi-particle 
of an itinerant electron and local spins, and accordingly
the magnetic polaron of FM cluster type is formed. 
As the clusters are far apart, electrons are trapped 
and their mobilities are reduced.
As the clusters are merged and the FM state begins to appear near $T_C$, 
the mobilities increase again.\cite{yu05,yu06}
That is, the insulator to metal transition is
caused by the percolation of the magnetic polarons.
It is thus evident that the peak and dip feature of the TEP 
in Fig.~\ref{fig1}(a) is closely related to the magnetic polarons
which are formed at $T_P$ and begin to be percolated at $T_D$.\cite{Tcd}

Near $T_C$, the TEP exhibits the strong magnetic field dependence
as seen in the magnetic and transport properties.\cite{bh07}
As show in Fig.~\ref{fig1}(b), the magnitude of the TEP 
diminishes with increasing the magnetic field and
its dip position shifts up in temperature.
Also notable is that the magnetothermoelectric power, $\Delta S=S(H)-S(0)$ is 
positive in this system.
This feature is also related to the formation of magnetic polarons. 
Because the thermal fluctuation of the local spins
is suppressed in the presence of the magnetic field, 
the magnetic polarons are easily formed and merged together. 
The percolation gives rise to the reduction of 
the electrical resistivity (the negative magnetoresistance) 
and the decline of the insulating phase. 
Then the emerging metallic phase suppresses
the peak and dip feature near $T_C$ in the TEP curve. 
Hence the TEP which has a negative value moves up with increasing 
the magnetic field.

%\section{Conclusion}

In conclusion, we have investigated the temperature dependence of the TEP
for the double exchange model in the dilute carrier concentration limit.
The TEP in the high temperature regime follows the Heikes formula 
independent on $J_H$ value.
The exchange coupling reduces the TEP in the intermediate temperature
regime. 
Near $T_C$, the TEP shows anomalous peak and dip feature and 
the positive magnetothermoelectric power, which are
explained by the concept of the magnetic polaron.

\begin{acknowledgments}
%\acknowledgments
This work was supported by the SRC/ERC program of MOST/KOSEF 
(R11-2000-071), the basic research program of 
KOSEF (R01-2006-000-10369-0), and by the POSTECH research fund.
\end{acknowledgments}


\begin{thebibliography}{99}
\bibitem{FS} The unexpected TEP behavior at extremely low temperature
        for $J_H = 0$ is thought to be caused by finite size of the system.
\bibitem{Tcd} In our results, $T_D$ is only slightly higher than $T_C$, and 
the conductivity variation is maximum near $T_C$. But, its magnitude begins to 
rise up rapidly near $T_D$.
\end{thebibliography}
\end{document}